# Characterization of the Chemical and Electrical Properties of defects at the Niobium-Silicon Interface

Cameron Kopas[1], Madhu Krishna Murthy[1], Christopher Gregory[1], Bryan Ibarra Mercado[1],
Daniel R. Queen[2], Brian Wagner[2], Nathan Newman[1]
[1]Materials Program, Arizona State University, Tempe, Arizona, 85287, USA
[2]Northrop Grumman Corporation, Mission Systems, Linthicum, Maryland 21090, USA

Abstract

The nature and concentration of defects near niobium-silicon interfaces prepared with different silicon surface treatments were characterized using current-voltage (I-V), deep level transient spectroscopy (DLTS), and secondary ion mass spectroscopy (SIMS). All samples have H, C, O, F, and Cl chemical contamination in the Si within 50 nm of the interface and electrically active defects with activation energies of 0.147, 0.194, 0.247, 0.339, and 0.556 eV above the valence band maximum ($E_{vbm}$). In all cases, the deep level defect concentration is dominated by the hole trap at $E_{vbm}$+0.556 eV, which we assign to a Nb point defect in Si, presumably $Nb_{Si}$. This defect is present with concentrations ranging from $7 \times 10^{13}$ to $5 \times 10^{14}$ cm$^{-3}$ and depends on the final surface clean process.

The optimum surface treatment used in this study is an HF etch followed by an in-situ 100 eV Ar-gas ion milling process. Higher energy ion milling is found to increase the electrically active Nb defect concentration in the Si, and increase the concentration of defects. The HF etch alone removes O from the interface, but results in significant H and F contamination, electrically-active point defect concentrations, and levels of Shockley-Reed-Hall recombination (i.e., Nb/Si Schottky diodes with an ideality factor, n, of ≈1.6). The RCA clean increases the depth and concentration of H, F, C, and Nb contamination.

## 1 Introduction

Superconducting electronic circuits and systems commonly use niobium because of its relatively high superconducting transition temperature ($T_C$=9.2 K) and relative ease in depositing high-quality thin-films. For applications that require ultra-low noise, including quantum computing, sensing, and information systems, undoped zone-refined Si substrates are most commonly used because of their wide availability and purity. Superconducting coplanar microwave resonators are an example of a device that uses this system at sub-Kelvin temperatures and single-photon levels to achieve minimal noise. However, in these devices, defect-induced microwave loss at the metal-dielectric interface is often the dominant "two-level-systems" noise source [1] [2] [3]. While surface engineering [4] [5] and removal of residue from microfabrication [6] can decrease this loss, the physical nature of the defects responsible for these losses is still unknown. In this investigation, we aim to identify the chemical and electrical properties of defects present at niobium-silicon interfaces and identify their physical nature when possible, using characterization techniques developed by the semiconductor industry.

In this study, time-of-flight secondary ion mass spectrometry (TOF-SIMS) has been used to measure the depth profile of contaminants and the level of intermixing at and near the Nb-Si interface. Rutherford Backscattering Spectrometry (RBS) and nuclear reaction analysis (NRA) are performed to determine layer thicknesses and concentrations. Current-voltage (I-V) and capacitance-voltage (C-V) profiling techniques identify changes in doping profile, barrier heights, the presence of Shockley-Reed-Hall recombination, an interfacial oxide layer, and surface leakage currents. Deep level transient spectroscopy (DLTS) further adds to these techniques, identifying concentrations, energy levels, capture cross-sections, and depth profiles of defects within the depletion region, and has the ability to determine the physical nature of the electronic defect by comparing to known electronic levels and capture cross-sections.

## 2 Experimental Parameters

$1 \times 1$ cm$^2$ p-type Si with 7–8 μm epitaxial B-doped ($8 \times 10^{16}$ cm$^{-3}$) on 500 μm B-doped $5 \times 10^{19}$ cm$^{-3}$ substrates were used. The top epitaxial Si layer is chosen with low enough doping concentration to make a low-leakage Schottky junction, while the base substrate is highly doped to reduce series resistance. Sequential substrate cleaning procedures were performed, and then the specimens were loaded into the vacuum deposition chamber within 180 seconds to minimize contamination and oxidation of the surfaces.

The "solvent only" cleaned substrates are immersed in acetone in an ultrasonic bath for 9 minutes, then in USP-grade anhydrous undenatured ethanol for 9 minutes. The "HF etch" sample processing follows the solvent clean with a 5-minute immersion in HF (2 vol% aqueous), after which we shake the HF off the sample without rinsing the surface to minimize organic contaminants. An ion milling process using a 5 cm Kaufman ion source at 60 eV Ar ion energy, 2 mA current for 2 minutes is then performed in-situ before the final Nb deposition for "ion mill" specimens. For the "RCA cleaned" specimens, we insert RCA clean "SC-1" and "SC-2" steps [7] before HF etch and ion milling; these consist of 10-minute immersion at 70 °C in NH$_4$OH + H$_2$O$_2$ to remove organic residue, followed by 10-minute immersion at 70 °C in HCl + H$_2$O$_2$ to remove metal contaminants and form an oxide layer. The RCA clean procedure is followed by the same 5-minute HF etch (2 vol% aqueous) to remove the oxide layer, then a 2-minute in-situ ion milling process with Ar ions, 60 eV, 2 mA before Nb deposition.

Niobium is deposited via magnetron sputtering in an un-baked cryo-pumped ultra-high vacuum (UHV) chamber with base pressure less than $2 \times 10^{-9}$ Torr. Source to substrate distance is ≈15 cm, with the Nb deposition rate ≈0.6 nm/second and a total film thickness of ≈200 nm. The device area is defined using a metal shadow mask to minimize confounding experimental effects such as etch damage or contamination from photolithographic processing. The device areas include $1 \times 1$ mm$^2$ squares for I-V and DLTS measurements, and a 2 mm diameter circle for the TOF-SIMS depth profile. This sample configuration enables us to meet experimental requirements for both chemical and electrical characterization on the same specimen to avoid run-to-run variation.

We prepare ohmic contacts on the back-side of the $5 \times 10^{19}$ cm$^{-3}$ boron-doped substrate immediately after removing the sample from vacuum by scratching the rough side of the film with a diamond scribe to break through the silicon oxide layer, then applying two pads of silver paint (SPI, Model Silver Conductive Paint, West Chester, PA) on an approximately $3 \times 8$ mm$^2$ area. The silver paint is then cured at room temperature overnight. Room-temperature I-V measurements across the two back-side contacts are performed to confirm the contacts are ohmic (i.e., linear and non-rectifying), and the large back-side contact area is used to ensure that the device capacitance is limited by the size of the $1 \times 1$ mm$^2$ Nb-Si Schottky barrier on the top surface.

TOF-SIMS depth profiling is performed with a Time of Flight Secondary Ion Mass Spectrometer (ION-TOF USA, Inc., Chestnut Ridge, NY). Negative secondary ions are monitored using a $^{69}$Ga primary analysis beam and Cs sputter beam at 1 kV, in an un-baked vacuum chamber with base pressure less than $2 \times 10^{-9}$ Torr. The relatively low 1 kV sputter energy minimizes beam intermixing and increases depth resolution; however, the low sputter rate (1.5 seconds/Å) increases the background signal from redeposition. TRIM [8] simulation estimates 2 nm of intermixing induced by the Cs beam at this energy, ignoring cascade effects from repeated sputtering; we estimate the experimentally-realized interface resolution of about 5 nm. SIMS sensitivity factors are only available for bulk Si [9], and since ionization potentials and charged-particle sputter yields can differ by orders of magnitude from those values in the highly intermixed interfacial Nb/Si region, all SIMS data in this volume are presented as secondary ion yield values normalized to the relative count rate of the host matrix (Nb or Si).

Rutherford Backscattering Spectrometry (RBS) and nuclear reaction analysis (NRA) use a 1.7 MV Tandetron accelerator (5.1 MeV maximum energy) with α-particles at 2.0 MeV (for RBS), 3.0 MeV for oxygen NRA, and 3.58 MeV for carbon NRA. Both RBS and NRA have a sensitivity of ≈0.01 at% in a thick layer, and a depth resolution of ≈0.5 nm.

In this investigation, we electrically characterize our devices using current-voltage (I-V) measurements using a picoammeter with an internal voltage source (Hewlett-Packard Model 4140B, Palo Alto, CA), and DLTS using a Semilab time-resolved capacitance meter (DLTS Model DLS-83, Budapest, Hungary). DLTS measurements were performed using a 20 μs filling pulse time, with rate windows ranging from 25–125 Hz, reverse bias of 1 V, and forward bias of +1.1 V to fill defects near the Nb/Si interface. A cryogenic cold stage refrigerator with a 50–350 K temperature range (Cryosystems Model LTS-21, Westerville, Ohio) cools the specimens for temperature profiling. For the DLTS measurements, the diodes are driven into reverse bias to empty traps, then a forward bias pulse fills traps, and the time-dependent change in capacitance is measured and related to the emission rate of the activated traps. We measure temperature spectrums at different rate windows (measurement frequency), generating frequency-temperature data for each peak with the intensity corresponding to the emission rate and concentration. An Arrhenius plot of these peak positions (T) and emission rates ($\tau_e$) is fit to the equation

$$\tau_e T^2 = \frac{\exp\left(E_T/k_B T\right)}{\gamma \sigma}$$

to obtain the energy level from the band ($E_T$) and capture cross section (σ) using the constant $\gamma=1.78\times10^{21}$ cm$^{-2}$s$^{-1}$K$^{-2}$ for p-type Si [10]. After identifying frequency-temperature pairs for specific energy levels, we perform depth profiles by setting the temperature and frequency to a constant value and then apply two excitation pulses with varying periods a set time apart to define the width of the space charge region. When possible, we compared measured activation energies and capture cross-section to those in the literature, to determine the physical origin of the defect, keeping in mind what chemical species are present in the region as measured by the TOF-SIMS measurements.

3   Results and Discussion
3.1   Chemical characterization of interfaces

The TOF-SIMS profiles in Figure 1 provide high depth and mass-resolved chemical profiles across the niobium-silicon interfaces. For this investigation, we focus primarily on the contaminants near the interface in the Si layer, since they are presumed to be responsible for the microwave losses [1]. In all specimens, TOF-SIMS identifies interfacial and near-interfacial contamination from hydrogen, carbon, oxygen, fluorine, and chlorine, likely remaining from substrate processing. TOF-SIMS reports only chemical contaminants and does not distinguish those that may be electrically inactive in Si. However, identifying chemical species can help identify sources of contamination, and help identify the origin of electrical defects measured with DLTS.

Because TOF-SIMS does not allow for easy quantification of interface contaminants, RBS and NRA were used to set an upper limit for interface contamination. In the "solvent only" specimen, we observe about 10 nm of $SiO_2$ at the interface, but all subsequent surface clean techniques remove $SiO_2$. For all other samples, there are not any contaminants observed in the Nb layer or the Si substrates to within the 0.01 at% detection limit of RBS. Since RBS and NRA have been established to be highly accurate and do not require a standard, it sets an upper limit to the level of impurities that can be present in the sample.

Figure 1 shows TOF-SIMS depth profiling near the Nb/Si interfaces, beginning 10 nm from the Si surface. We detect H, C, O, F, and Cl in all specimens. In the "Solvent Only" specimen the H concentration drops off near the interface, with a 2$^{nd}$ peak about 25 nm from the interface and reaching a minimum ≈60 nm into the Si, while the F, O, C, and Cl reach their background yields within 20–30 nm. After performing an HF Etch, intended to remove surface $SiO_2$, we see a decrease in oxygen at the interface, as expected, but with additional C, H, and F in the silicon, indicating some contamination or residue from the HF clean process. Following an HF etch, the addition of the 60 eV in-situ ion mill ("HF & Ion Mill") decreases interfacial H and F, with Cl and F peaks concentrated on the Nb-side of the interface, resulting in the least contaminated sample in this series. In the RCA & HF & Ion Mill specimen, the width of the Nb/Si interface broadens by about 5 nm, there is increased C in the Si, and most significantly, there is a 2$^{nd}$ peak of F concentration about 20 nm into the Si. The 2$^{nd}$ peak suggests that the F is migrating into the substrate instead of passivating a flat Si surface, either due to roughness or reaction with the substrate. From these data, we conclude that the interface with the lowest measured levels of contamination is produced by an HF etch followed by an in-situ ion mill at 60 eV. An RCA etch can introduce additional contamination to the Nb/Si interface that is not removed by the HF etch or ion mill.

Based on the observations that surface impurity concentrations are higher on the Nb-side of the interface, it is likely that some mobile contaminants on the surface preferentially accumulate in the Nb layer due to a gettering effect. Gettering on the metal side can be beneficial to devices which rely on a clean dielectric material (assuming the "dirty" metal near the interface does not cause a large drop in the superconducting order parameter), and has been used to remove defects such as iron from Si wafers [11]. This effect may be further enhanced by annealing or the introduction of other materials such as Ta and Al [12]. In addition to gettering, the concentration of electrically active defects can be reduced by annealing Nb/Si structures, as many simple defects anneal out of silicon at temperatures between 200 and 400 °C [13].

3.2   Electrical characterization – I-V and DLTS

Current-voltage characteristics of a Schottky diode are sensitive to small changes in the atomic structure and contamination at the interface [14], where electronically active defects can increase the ideality factor (n) by acting as recombination centers and altering the barrier height ($\phi_B$) by contributing to interface trapped charge. We fit the current in the forward bias to the thermionic emission equation

$$I = SA^*T^2 e^{-\phi_B/V_t} e^{(V-IR)/nV_t}$$

where S is the area of the device, A* is the Richardson constant (8.1 A/K$^2$cm$^2$ for Si), T is the measurement temperature, $\phi_B$ is the effective barrier height, $V_t$ the thermal voltage ($k_BT$), R the series resistance, and n is the ideality factor. When n=1 the transport is described by thermionic emission while n>1 corresponds to defect assisted Shockley-Reed-Hall recombination, surface leakage currents, or an interfacial oxide layer. Even in the most-ideal diodes made, the ideality factors fall slightly above 1, typically 1.02–1.05, as a result of barrier-lowering effects at higher voltages from image force and tunneling [15]. As is commonly encountered, the defect levels fall between the electron and hole quasi-Fermi levels in forward bias, and then the Shockley, Reed, Hall recombination results in an ideality factor of ≈2. In the case of diodes with pinholes or surface conduction, the lowest barrier will dominate until the current density of the path saturates from the local spreading resistance, and the net current acts as multiple diodes in parallel, as discussed in ref [15]. Leakage from peripheral surface conduction is often linear in voltage and can be removed by the application of a mesa etch. We did not find strong evidence of such peripheral surface conduction in our devices.

Figure 2 shows representative I-V scans of a single Nb/Si Schottky diode at temperatures ranging from 50-300 K. At high temperatures, series resistance dominates the device properties at relatively low voltages, while at low temperatures, the recombination currents dominate. For these reasons, we chose to focus our analysis on the I-V properties measured at 200 K.

Figure 3 shows I-V characteristics for each of the Si-surface cleaning procedures along with the ideality factor and barrier height values from our fit. The data are expressed in a semi-log plot to visualize the slope of the graph being inversely proportional to the ideality factor. An ideality factor of 1 corresponds to ≈60 mV per decade. The "most ideal" (i.e., with the lowest ideality factor of n=1.17) occurs in Nb/Si samples prepared with an HF etch and in-situ ion mill, followed by the specimen which received only an HF etch (n=1.6), while the "solvent only" and RCA etched samples exhibit n>2. Nb/Si samples that have undergone the RCA clean have a higher ideality factor than the other surface preparation methods produce. Figure 4 shows representative DLTS temperature spectra for the specimen cleaned with an HF etch and 60 eV ion mill, and the associated Arrhenius plot is in Figure 5. All specimens display similar DLTS spectra, dominated by a negative peak near $T_{peak}$ = 250K, followed by four additional defects at lower temperatures and much lower intensity. All resolved defect peaks are negative, corresponding to majority carrier hole traps.

Table 1 summarizes the fitted activation energies and capture cross sections observed in every specimen with their possible identifications, while Table 2 compares the defect concentrations of these defects between surface treatments. Defect activation energies will be referred to by the common shorthand notation H (for hole) followed by parentheses with the energy above $E_{vbm}$ in eV. The dominant defect H(0.556) (at $T_{peak}$=250 K)–which refers to a hole trap at $E_{vbm}$+0.556 eV—is most likely associated with a niobium defect in p-type silicon, as reported in [16]. Interestingly, this defect changes activation energy and capture cross section before and after the addition of an HF etch. This may correspond to the presence of a nearby "spectator" defect or other forms of a change in environment. In the "solvent only" sample, the energy of this peak is $E_V$+0.447 eV and its capture cross section $\sigma=2.62\times10^{-15}$ cm$^2$, while all subsequent surface treatments this defect occurs at $E_V$+0.556 eV with $\sigma=3.72\times10^{-13}$ cm$^2$, but both activation energies correspond to reported energy levels for niobium in p-type silicon [16] [17]. The large ≈$10^{-13}$ cm$^2$ capture cross section we observe for H(0.556) suggests its origin may be associated with a defect complex involving niobium, but present literature reports do not speculate on the exact nature of this defect. The H(0.194) defect matches energy levels for a silicon divacancy with a +1 charge, $(V_{Si}-V_{Si})^+$ [13], and also matches reported energy levels associated with F and Cl in Si [17]. The H(0.339) defect matches the energy for a thermally-activated $V_{Si}^{2+}$ migration, and a $V_{Si}^0$ migration [13], and is close to both the energy and capture cross section reported for a $C_s$-$C_i$ pair [18].

However, vacancies in silicon are highly mobile and expected to anneal out at temperatures above 200 K [13], so these are unlikely to survive except if trapped by another defect. While the listed origins represent some possibilities, confidently determining the physical origin of energy levels requires further investigations.

The ideality factor inferred from the I-V characteristics correlates reasonably well with the total defect concentration measured by DLTS, as shown in Table 2. This is consistent with expectations from the Shockley-Read-Hall recombination theory [19] [20]; however, we are not aware of any prior work that has documented this important relationship in niobium-silicon diodes. In terms of both chemical and electrical properties, the HF etch and ion milled Si surface yields diodes with the lowest contamination levels and best electrical properties found among the different cleaning methods.

### 3.3 Optimization of Ion Mill Energy

This section compares the interfaces of Nb on p-type Si, where the Si was cleaned with an HF etch followed by an Ar ion mill at 60, 100, or 150 eV, with the same total ion dose. We estimate that ion milling at 60, 100, and 150 eV removed 0.4, 0.8, and 4 nm of Si, respectively, from the Si surface [21]. Higher energy incident ions are expected to remove more surface atoms, and more interface contamination, but with a higher probability of introducing Si vacancies, interstitials, and other native defects.

Figure 6 shows TOF-SIMS depth profiles for Si cleaned with an HF etch followed by in-situ ion milling at 60, 100, and 150 eV. Interestingly, we do not see any change in Nb/Si intermixing as a result of higher ion mill energy (within the approximately 5 nm resolution for these sputter depth profiles). The nature and level of chemical contamination observed in the Si layer does not exhibit significant differences between the three samples.

I-V measurements performed at 200 K are shown in Figure 7. Increasing the ion mill energy from 60 eV to 100 eV reduces the current in reverse bias by almost 2 orders of magnitude, and decreases the ideality factor from n=1.17 to n=1.07. Increasing the ion mill energy to 150 eV results in nearly the same curve as the 100 eV-cleaned specimen, with slightly improved ideality factor n=1.06. These results indicate that ion milling with energy greater than 100 eV can improve the Schottky barrier properties of the Nb/Si interface.

Table 2 includes the defect concentrations for each activation energy determined by DLTS for the HF and ion milled Schottky diodes. The concentration of the dominant defect, H(0.556), is lowest in the specimen ion milled with 100 eV Ar ions, approximately half the concentration observed in the 60 eV and 150 eV ion milled specimens. The increase in the concentration of H(0.339) and H(0.247) with higher energy ion milling indicates that these are ion-induced defects. The 100 eV ion milled specimen has the lowest total deep level defect concentration of any measurements, with $N<10^{14}$ cm$^{-3}$. These data confirm that surface cleaning procedures must be optimized to minimize the total defect concentration near the interface without introducing additional defects from substrate cleaning processes.

DLTS was used to perform depth profiles of electrically active defects, shown in Figure 8. Using DLTS depth profiling techniques, we can probe within ≈10 nm of the interface. We find that the majority of defects are concentrated within the first 50 nm from the interface, matching the range of interface contamination observed with TOF-SIMS. The defects H(0.147) and H(0.339) have nearly constant concentration across the measurement range. Two defects, H(0.247) and H(0.556), exhibit a second peak in concentration at approximately 210 and 240 nm deep, respectively. This deep accumulation near the edge of the depletion region suggests that these are mobile defects and have some charge that drives them into the substrate.

## 4  Conclusions

We have used TOF-SIMS, I-V, and DLTS techniques to compare the chemical and electrical properties of defects at the Nb-Si interface. We identify chemical contamination at the interface from H, C, O, F, and Cl, within the top 50 nm, and deep level defects at $E_{vbm}+0.147$, $E_{vbm}+0.194$, $E_{vbm}+0.247$, $E_{vbm}+0.339$, and $E_{vbm}+0.556$ eV. Based on DLTS depth profiling, most electronic defects are within the 50 nm closest to the Nb/Si interface, with the defects at $E_{vbm}+0.556$ eV and $E_{vbm}+0.247$ exhibiting a second, albeit much lower concentration, peak of charged defects 200–250 nm deep. Comparing different surface preparation techniques shows a minimum total defect concentration when the Si surface is cleaned with an HF etch and a 100 eV in-situ Ar ion mill. More aggressive chemical surface treatments (RCA clean) and higher energy ion milling (i.e., Ar at 150 eV) result in significantly more defects at the interface.

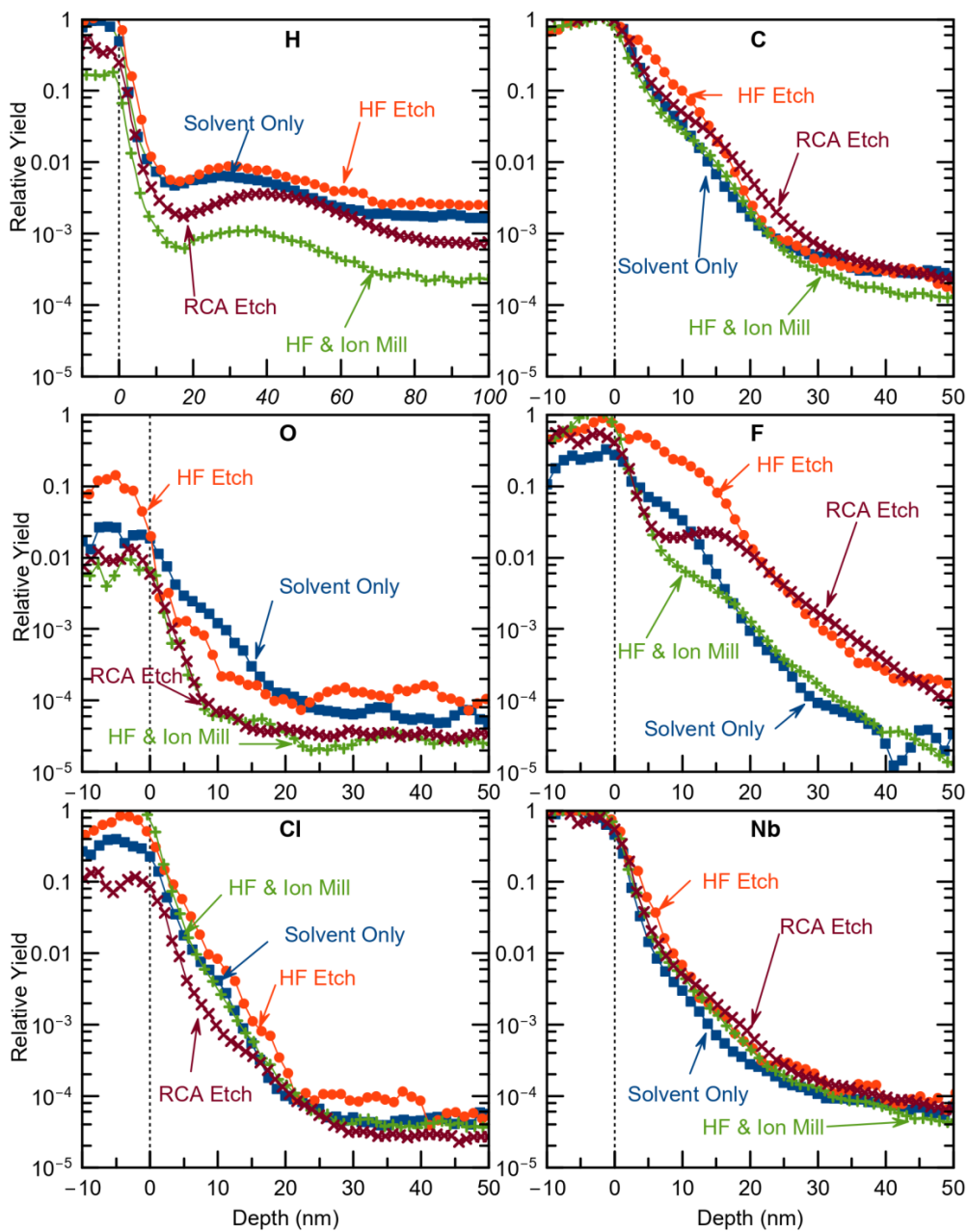

Figure 1: TOF-SIMS depth profiles comparing Nb/Si interface contaminants between Si surface preparation experiments. Each depth profile is performed using a 1 kV Cs sputter, $Ga^+$ analysis beam, and collecting negative secondary ions. Yield is relative to the sum of Nb and Si counts to minimize matrix effects. The depth where Nb and Si yields are equal define the x-axis zero for the depth. Hydrogen is displayed to 100 nm in depth, while all other elements are shown from -10 to 50 nm. The HF & Ion Mill specimen has the lowest interface impurity density in the Si.

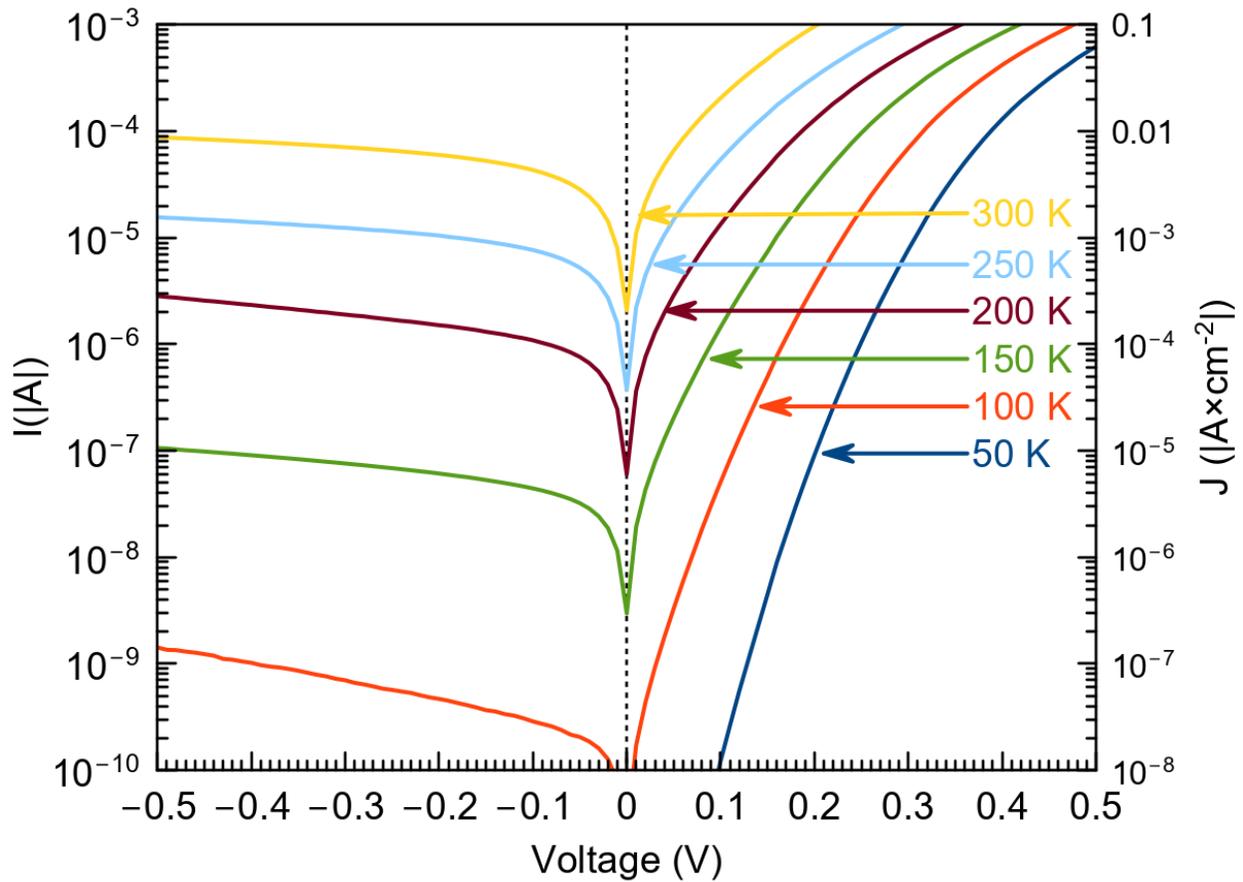

Figure 2: Representative I-V plot for an Nb/Si Schottky diode measured at different temperatures. This Si surface was cleaned with solvent only; its ideality factor at 200 K is n=2.3, indicating the dominance of Shockley-Reed-Hall recombination and/or the presence of an interfacial oxide layer.

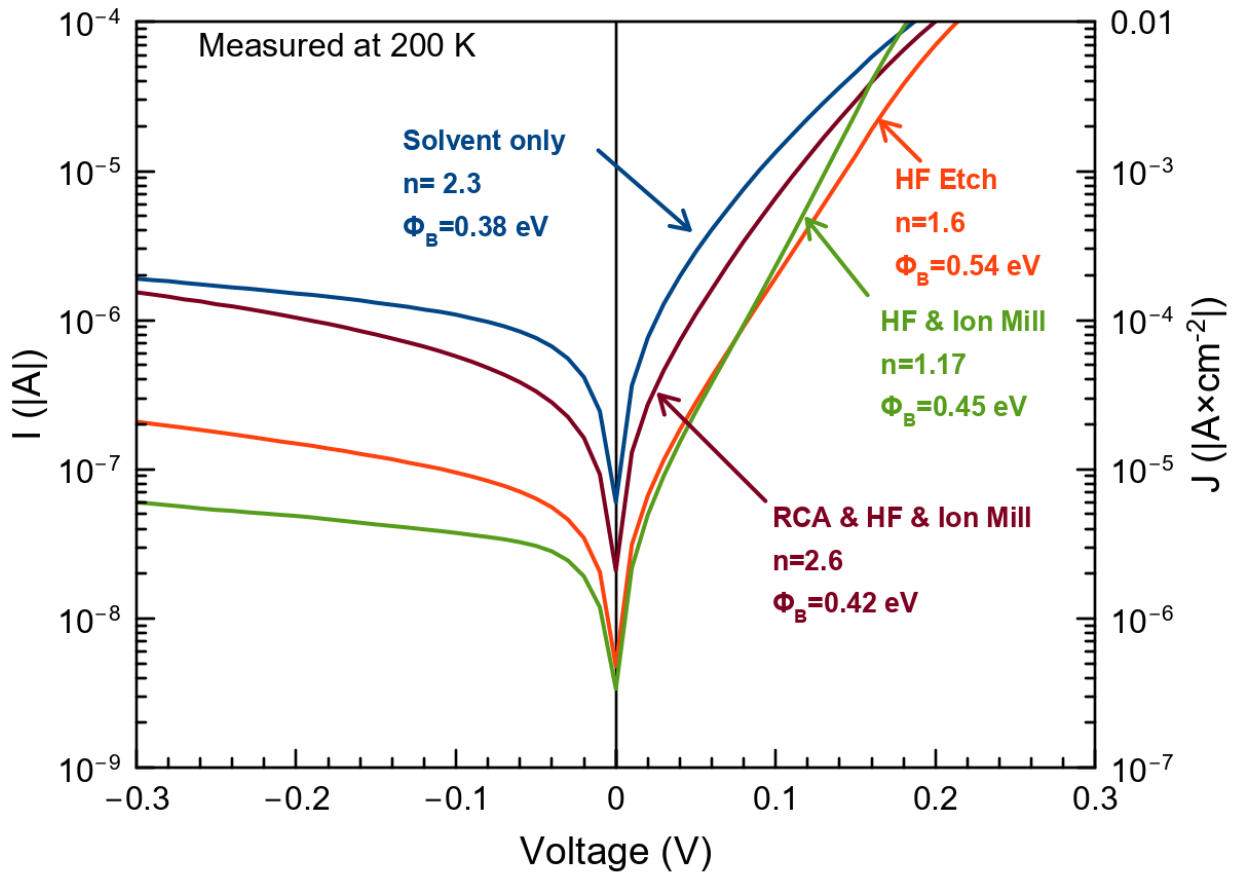

Figure 3 I-V measurements performed at 200 K comparing Nb/Si structures prepared with different Si surface treatments. Curves are labeled with the Si surface treatment performed before Nb deposition and their corresponding ideality factor and barrier height. Each ion mill energy in this figure is Ar≈60 eV. The curves are: "solvent only" (n=2.3, $\phi_B$=0.38 eV), "HF Etch" (n=1.6, $\phi_B$=0.54 eV), "HF Etch and Ion Mill" (n= 1.17, $\phi_B$=0.45 eV), and "RCA & HF & Ion Mill" (n=2.6, $\phi_B$=0.42 eV)

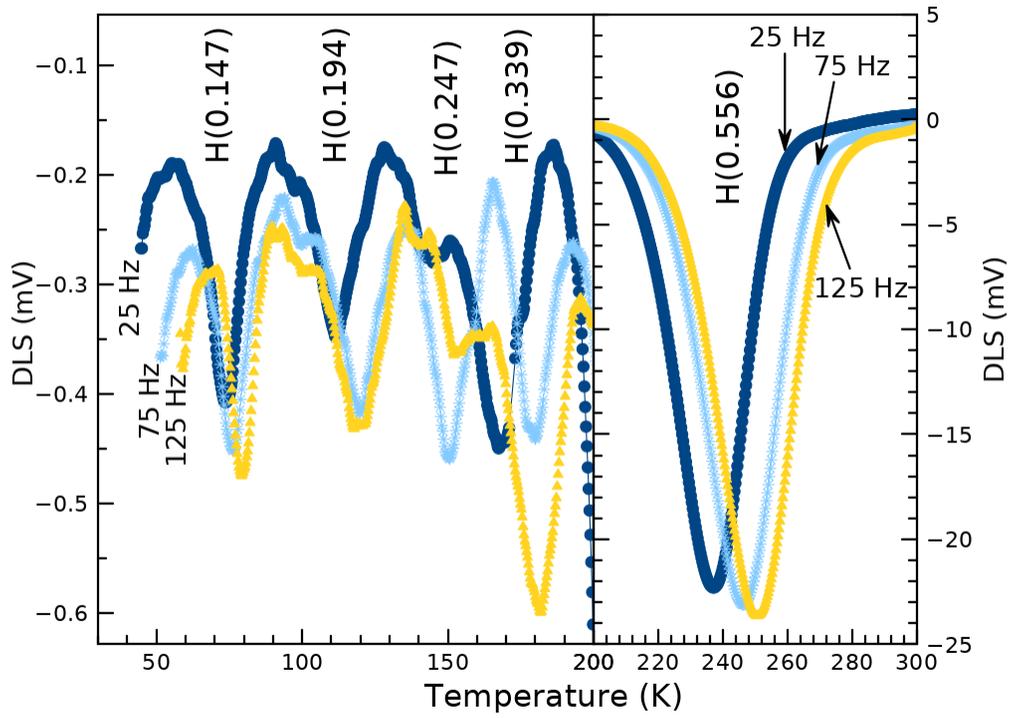

Figure 4: Representative DLTS spectra performed at 25, 75, and 125 Hz for a Nb/Si diode prepared with HF etch and 60 eV ion mill. The spectra are dominated by the peak identified as H(0.556) (right), attributed to a Nb defect [16] [17].

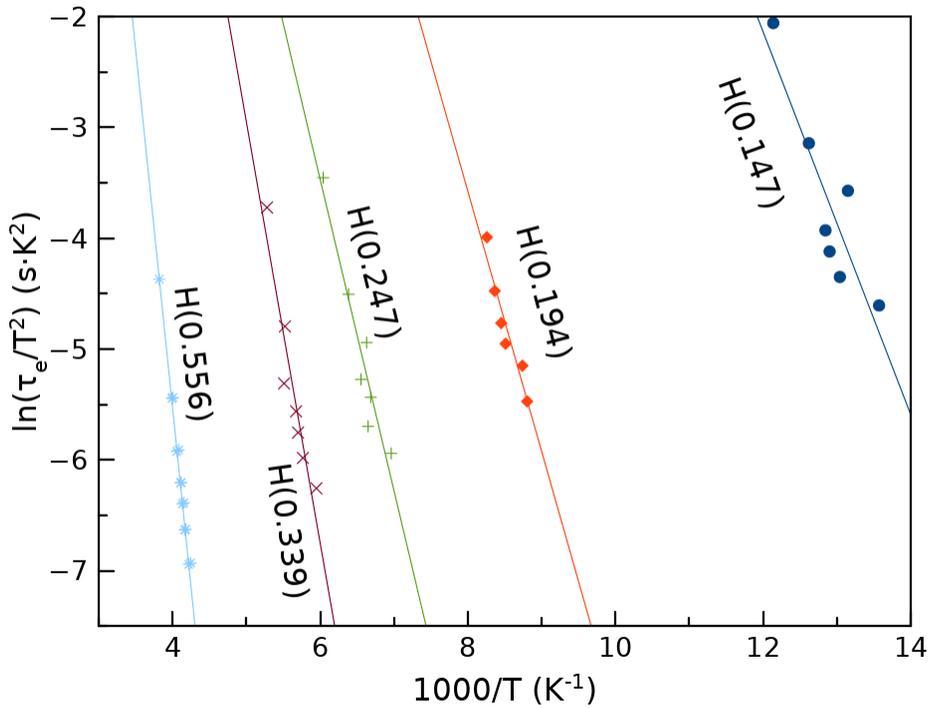

Figure 5: Representative Arrhenius plot of the DLTS peaks for the diode prepared with a HF and 60 eV ion milled surface clean measured with rate windows 25, 35, 45, 55, 75, 125, and 400 Hz. Each point corresponds to the temperature and emission rate of a peak measured in DLTS-temperature spectra at different frequencies. Linear least-mean-square error fits are drawn through points corresponding to the same peak measured at different frequencies. The slope of each line corresponds to the energy level of the trap, and the capture cross section is inferred from the y-intercept.

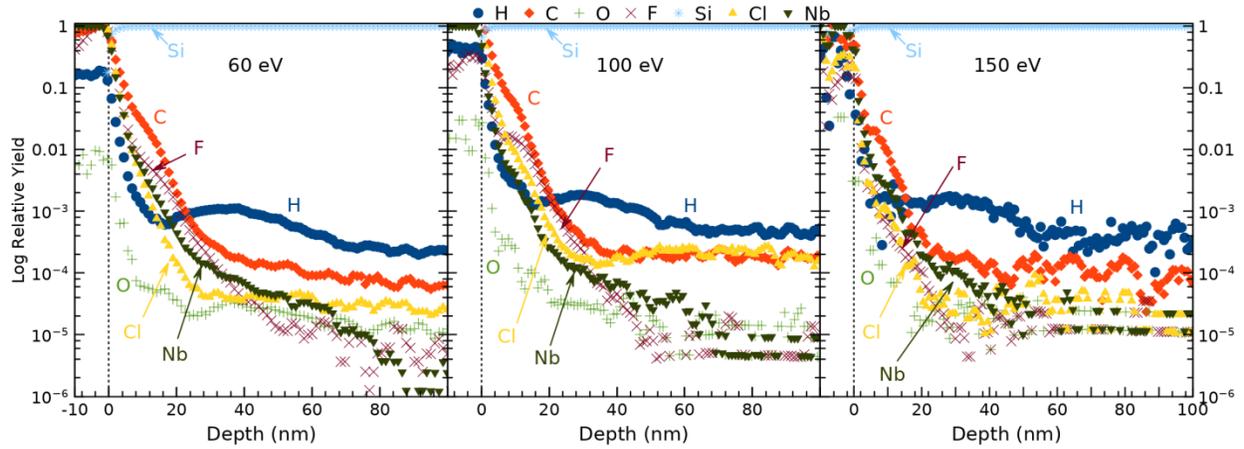

Figure 6: TOF-SIMS depth profiles showing contaminants measured near the Nb/Si interface for Si substrates cleaned with HF etch followed by a 2-minute in-situ ion mill at **(a)** 60 eV, **(b)** 100 eV, and **(c)** 150 eV. The zero of the x-axis scale is set at the depth where the Nb and Si yields are equal.

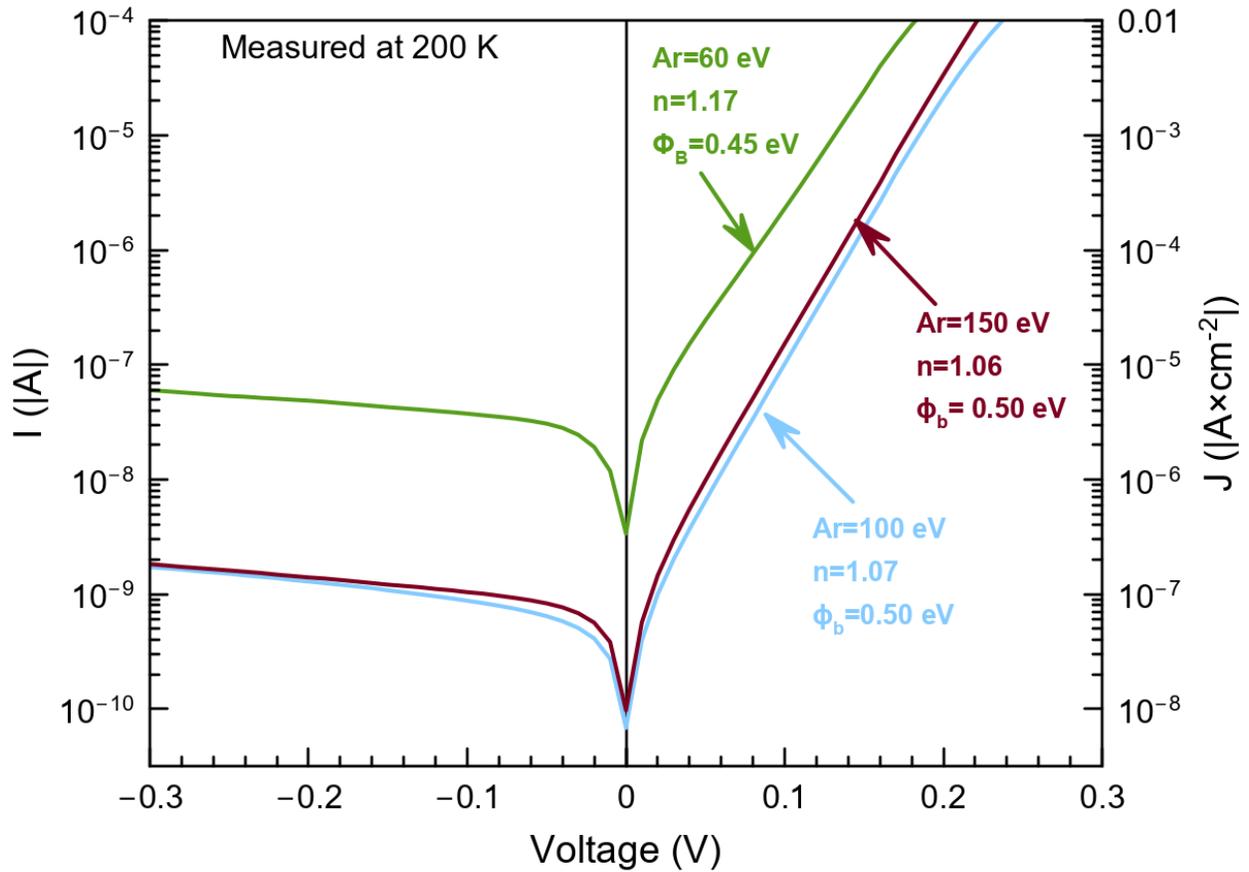

Figure 7: I-V measurements performed at 200 K comparing Nb/Si structures prepared with an HF etch followed by an in-situ Ar ion mill, with varying mill energies. Curves are labeled with the energy of incident Ar ions and their corresponding ideality factor and barrier height. They are: "Ar=60 eV" (n=1.17, $\phi_B$=0.45 eV), "Ar=100 eV" (n=1.07, $\phi_B$=0.50 eV), and "Ar=150 eV" (n=1.06, $\phi_B$=0.50 eV).

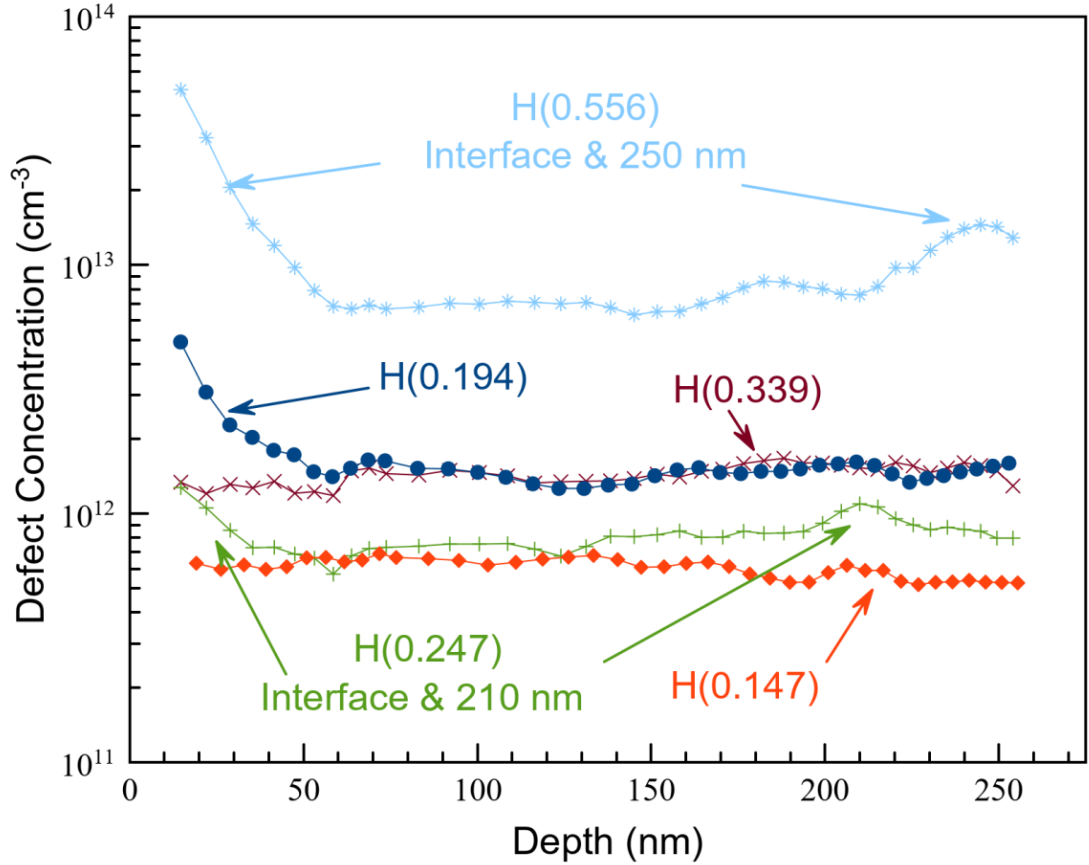

Figure 8: Representative defect depth profiles for HF etched and 150 eV ion milled specimen. Defects are primarily concentrated at the interface, within the first 50 nm, while the DLTS measurements of defects H(0.556) and H(0.247) have a second peak between 200 and 250 nm.

| Designation | Energy Level (meV) | $\sigma$ (cm$^2$) | T$^a_{peak}$ (K) | Possible origins |
|---|---|---|---|---|
| H(0.147) | E$_v$+147 ($\pm$2) | 6.7×10$^{-14}$ ($\pm$0.3×10$^{-14}$) | 75 | – |
| H(0.194) | E$_v$+194 ($\pm$7) | 6.2×10$^{-15}$ ($\pm$5.8×10$^{-14}$) | 120 | (V$_{Si}$-V$_{Si}$)$^+$ [13], F [17], Cl [17] |
| H(0.247) | E$_v$+247 ($\pm$5) | 1.4×10$^{-15}$ ($\pm$1.5×10$^{-15}$) | 150 | – |
| H(0.339) | E$_v$+339 ($\pm$15) | 2.1×10$^{-14}$ ($\pm$1.3×10$^{-14}$) | 180 | C$_s$-C$_i$ pair [18], V$_{Si}^{2+}$, or V$_{Si}^0$ migration [13] |
| H(0.556)$^b$ | E$_v$+556 ($\pm$12) | 3.72×10$^{-13}$ ($\pm$1.5×10$^{-13}$) | 250 | Nb [16] [17] |

Table 1: Average defect energy levels, their capture cross sections, temperatures, and potential defect identifications. All defects measured are majority carriers (hole traps). The assignment of possible origins comes from using both the information on the potential species in the area obtained from the SIMS measurements and from comparing energy level and capture cross section to the literature. Reported capture cross sections are calculated from Arrhenius plots, and may have large error.
$^a$ Nominal peak temperature with the rate window set to 75 Hz.
$^b$ The values for the defect at T$_{peak}$=250 K in the minimally-processed "solvent only" sample deviate from the others and are not included in these averages. Its values are E=E$_v$+447 meV, and $\sigma$=2.62×10$^{-15}$ cm$^2$.

|  | Ideality Factor | H(.147) (cm$^{-3}$) | H(0.194) (cm$^{-3}$) | H(0.247) (cm$^{-3}$) | H(0.339) (cm$^{-3}$) | H(0.556) (cm$^{-3}$) |
|---|---|---|---|---|---|---|
|  |  |  | $(V_{Si}^{\cdot}\text{-}V_{Si})^{+}$ [13], F [17], Cl [17] |  | $C_s\text{-}C_i$ pair [18], $V_{Si}^{2+}$, or $V_{Si}^{0}$ migration [13] | Nb [16] [17] |
| Solvent only | 2.3 | *>7.4×10$^{12}$ | *>6.4×10$^{12}$ | *>6.3×10$^{12}$ | 7.2×10$^{12}$ | 2.50×10$^{14}$ |
| HF Etch | 1.6 | *>6.5×10$^{12}$ | *>1.3×10$^{12}$ | *>2.4×10$^{12}$ | 6.6×10$^{12}$ | 1.73×10$^{14}$ |
| RCA & HF & Ion Mill | 2.6 | 1.1×10$^{13}$ | 6.7×10$^{12}$ | 6.4×10$^{12}$ | 6.6×10$^{12}$ | 4.86×10$^{14}$ |
| HF & Ion Mill (60 eV) | 1.17 | 1.9×10$^{12}$ | 3.1×10$^{12}$ | 1.3×10$^{12}$ | 2.8×10$^{12}$ | 1.66×10$^{14}$ |
| HF & Ion Mill (100 eV) | 1.07 | *>1.92×10$^{12}$ | *>2.72×10$^{12}$ | 8.1×10$^{12}$ | 1.1×10$^{13}$ | 7.4×10$^{13}$ |
| HF & Ion Mill (150 eV) | 1.06 | 1.4×10$^{13}$ | 3.0×10$^{13}$ | 1.8×10$^{13}$ | 3.1×10$^{13}$ | 1.96×10$^{14}$ |

Table 2: Comparison of defect concentrations determined by DLTS, and ideality factor determined by I-V for each specimen. All values are minimum values. The symbol *> refers to lower-bound values. In these cases, the peaks are not well defined as a result of the small signal being on the order of the measurement noise.